\documentclass[12pt,english,a4paper]{article}
\usepackage{babel}
\usepackage{amsmath,amsfonts,dsfont}
\usepackage{graphicx}
   
\def\fslash#1{\setbox0=\hbox{$#1$}#1\hskip-\wd0\hbox to\wd0{\hss\sl/\/\hss}}

\newcommand{\eg}{\emph{e.g.\ }}
\newcommand{\ie}{\emph{i.e.\ }}

\newcommand{\unitmatrix}{\mathds{1}}
\newcommand{\ket}[1]{|\!#1 \rangle}
\newcommand{\bra}[1]{\langle #1\!|}
\newcommand{\braket}[2]{\langle #1 | #1 \rangle}
\newcommand{\anti}[1]{$\overline{\text{D8}}$}
\newcommand{\bea}{\begin{eqnarray}}
\newcommand{\eea}{\end{eqnarray}}
\newcommand{\nn}{\nonumber}

\title{A note on fermions in holographic QCD}
\author{Rainer Heise and Harald G Svendsen}
\date{\today}

\begin{document}
\begin{flushright}
AEI-2007-010
\end{flushright}

\bigskip

\begin{center} {\Large \bf A note on fermionic mesons in  }
\bigskip
{\Large \bf holographic QCD }
\end{center}
 
\bigskip\bigskip\bigskip
 
\centerline{\bf Rainer Heise and Harald G. Svendsen}

\bigskip\bigskip
\centerline{\it Max-Planck-Institut f\"ur Gravitationsphysik}
\centerline{\it (Albert-Einstein-Institut)}
\centerline{\it D-14424 Potsdam, Germany}

\bigskip
\centerline{\small \tt rainer.heise, harald.svendsen}
\centerline{\small \tt @aei.mpg.de}

\bigskip\bigskip

\abstract
We study the fermionic sector of a probe D8-brane in the supergravity
background made of D4-branes compactified on a circle with
supersymmetry broken explicitly by the boundary conditions. 
At low energies the dual field theory is effectively four-dimensional
and has proved surprisingly successful in recovering qualitative and
quantitative properties of QCD.
We investigate fluctuations of the fermionic fields on the probe D8-brane 
and interpret these as mesinos (fermionic superpartners of mesons). 
We demonstrate that the masses of these modes are comparable to meson masses 
and show that their interactions with ordinary mesons are not suppressed. 
\thispagestyle{empty}
\newpage
\setcounter{page}{1}
\section{Introduction}

Since the discovery of the AdS/CFT correspondence~\cite{Maldacena:1997re, Witten:1998qj, Gubser:1998bc}, there has
been a lot of interest in trying to find a gravity dual of 
QCD.
Important progress was made with the realisation~\cite{Karch:2002sh} that flavour degrees of freedom can be incorporated by adding $N_f$ D-branes in the probe approximation where their back-reaction is ignored. 
This is an approximation, which is valid only if the number of flavours is much less than the number of colours, $N_f\ll N_c$.

Chiral symmetry breaking in holography was studied in a D3/D7-brane model in \cite{Babington:2003vm}.
An interesting D4/D6-brane model was presented in ref.~\cite{Kruczenski:2003uq} where $N_c$ D4-branes compactified on a circle make a supergravity background, in which $N_f$ flavour branes are put in as probes. 
On scales smaller than the Kaluza--Klein scale associated with the compact direction, the dual gauge theory is effectively four-dimensional and exhibits many of the qualitative features of QCD.

Based on this model, Sakai and Sugimoto presented a similar
D4/D8-brane model \cite{Sakai:2004cn,Sakai:2005yt}, which comes surprisingly close to recover the meson spectrum as found experimentally.
In particular, the spectrum contains a massless pion, associated with a spontaneously broken chiral symmetry.
Moreover, the effective theory for the pions was found to contain the Skyrme model.
It has been argued that baryons can be constructed in the supergravity description by introducing a baryon vertex~\cite{Witten:1998zw}, which in the D4/D8 context is a D4-brane wrapped on a four-sphere.
This object is interpreted as the Skyrmion on the probe D8-brane.
Several investigations of the baryons from this point of view have been performed recently in \eg refs.~\cite{Hata:2007mb, Hong:2006ta, Hong:2007ay, Hong:2007kx, Nawa:2006gv, Nawa:2007gh, Nitta:2007zv}. 

As a holographic description of QCD there are some fundamental difficulties with the D4/D8 model. 
One problem is that on scales small compared to the radius of the compact direction (\ie in the UV) the gauge theory is inherently five-dimensional. 
Moreover, the mass scale of the scalar and vector mesons is the same as the mass scale $M_{kk}$ of the Kaluza--Klein modes arising from the compact direction.
That is, the model predicts a tower of non-observed Kaluza--Klein modes with similar masses to the observed mesons.
Another problem is that the D4/D8-brane configuration is
non-supersymmetric and unstable, the unstable modes being the scalar fields corresponding to the radius of the circle and to the asymptotic brane-anti-brane separation. 
However, in the supergravity limit these modes become non-normalisable and as such cause no worry.

The addressed problems in the D4/D8-brane holographic model of QCD might well be related to the fact that we are only working in the probe approximation. 
Ideally, the flavour D8-brane back-reaction should be taken into account, but how to do this is only partly understood at the moment. 
For some attempts to go beyond the probe approximation, see \eg refs.~\cite{Erdmenger:2004dk, Burrington:2004id, Casero:2006pt, Benini:2006hh, Benini:2007gx}. 

In this paper, we investigate an implication to the Sakai--Sugimoto model of string theory having a supersymmetric spectrum. 
Although the supersymmetry of the D4 background is explicitly broken by the boundary conditions, there is still a supersymmetric spectrum. 
In particular, the D8-brane probe action has a fermionic part in addition to the bosonic Dirac-Born-Infeld and Chern-Simons parts. 
The fluctuations of these fermionic modes should be interpreted as the supersymmetric partners of the mesons, so we refer to them as \emph{mesinos}.
Being aware of their presence in the full spectrum, 
a natural question is therefore what their masses are.
Since mesinos are not found experimentally, nor in the QCD, 
these masses ought to be very large.
However, due to the absence of different scales it seems unlikely that this is so. 
A priori, we therefore expect mesino masses to be
comparable to the meson masses, of order $M_{kk}$.
We compute these masses explicitly and conclude that the
expectation is true: Mesinos appear on the same scale as mesons.

The question then arises how this ``unwanted'' (from the
phenomenological point of view) mesino sector affects the ``good'' meson sector. 
To get a partial answer to this question, we investigate
mesino interaction terms, in particular, we compute trilinear
meson--mesino--mesino couplings. 
One could be tempted to speculate that these interactions are somehow suppressed and that the mesino sector therefore would not interact with the meson sector. 
However, we shall find that this is not the case. 
The mesinos do indeed interact with the mesons. 

The paper is organised as follows. 
Section~1 contains a review of the Sakai--Sugimoto model, introducing 
the notation used throughout the paper. 
Moreover, the spectrum of the mesons in this model is re-derived.
Section~2 displays the computation of the spectrum of the fermionic 
fluctuations of the D8-brane, which are identified as mesinos.
Section~3 describes how these mesinos interact with bosonic fields. 
The effective four-dimensional interaction vertices are given explicitly.
Section~4 is dedicated to a discussion of our results and concludes the paper.

\subsection{The Sakai--Sugimoto Model}

The Sakai--Sugimoto model~\cite{Sakai:2004cn} is a D4/D8-brane
configuration with $N_c$ D4-branes wrapping a circle 
and $N_f$ flavour D8-\anti{D8}-branes. 
Schematically, the brane configuration is as follows:
\begin{center}
\begin{tabular}{lcccccccccc}
&0&1&2&3&4&5&6&7&8&9
\\
D4 &*&*&*&*&*
\\
D8-\anti{D8} &*&*&*&*&&*&*&*&*&*
\end{tabular}
\end{center}
The 4-direction is the circle, and anti-periodic boundary conditions
for the fermions on this circle break the supersymmetry explicitly.

This D4-brane supergravity background is given by~\cite{Witten:1998zw}
\begin{equation}
\begin{split}
  & ds^2 = \left(\frac{U}{R}\right)^\frac{3}{2} \left(
  \eta_{\mu\nu}dx^\mu dx^\nu + f(U)d\tau^2 \right)
  +\left(\frac{R}{U}\right)^\frac{3}{2} \left( \frac{dU^2}{f(U)}+U^2
  d\Omega_4^2 \right),
\\
  & e^{\phi} = g_s \left(\frac{U}{R}\right)^{\frac{3}{4}},
  \qquad
  F_4=dC_3=
      \frac{3 R^3}{g_s}\epsilon_4,
  \qquad
  f(U)=1-\frac{U_{kk}^3}{U^3},
\end{split}
\end{equation}
where $\mu,\nu=0,\dots,3$ and $\tau$ is the compact direction, which
the D4-branes wrap. $V_4=\frac{8\pi^2}{3}$ is the volume, 
$\epsilon_4$ is the volume form and $d\Omega_4^2$ is the line element
of a unit $S^4$, respectively. 
$R$ and $U_{kk}$ are constant parameters and $R^3=\pi g_s N_c l_s^3$
where $g_s$ is the string coupling, $l_s\equiv\sqrt{\alpha'}$ is the string length and
$N_c$ is the number of D4-branes.
The coordinate $U$ 
satisfies $U\ge U_{kk}$, and to avoid a conical singularity, the
coordinate $\tau$ must have a period 
\begin{equation}
  \delta\tau=\frac{4\pi}{3} R^\frac{3}{2} U_{kk}^{-\frac{1}{2}}
  \equiv \frac{2\pi}{M_{kk}},
  \qquad
  M_{kk}=\frac{3}{2}U_{kk}^\frac{1}{2} R^{-\frac{3}{2}},
\end{equation}
where we have defined the Kaluza--Klein mass $M_{kk}$. Below this
energy scale, the field theory is effectively four-dimensional.
In the limit $U_{kk}\to 0$, the $\tau$ direction uncompactifies and
we recover the background built out of flat D4-branes.

We are going to study a particular embedding of the
D8-\anti{D8}-branes, in which the D8 and \anti{D8} are antipodal in the
$(U,\tau$) coordinates and smoothly join together at $U=U_{kk}$.
For this purpose, it is useful to make the
change of coordinates $(U,\tau) \to (y,z)$, defined through
\begin{equation}
   U^3 = U_{kk}^3(1+\frac{r^2}{U_{kk}^2}),
  \quad
   \theta =\frac{2\pi}{\delta\tau}\tau,
\end{equation}
\begin{equation}
  y = r \cos\theta,
  \quad
  z = r \sin\theta.
\end{equation}
In the $(y,z)$ coordinates, the probe D8-\anti{D8}-branes are extended
along the $z$ direction and located at $y=0$. This configuration
corresponds to massless quarks in the gauge field theory and was
demonstrated in the original paper~\cite{Sakai:2004cn} to be a stable
solution of the probe action.

\subsection{Meson spectroscopy}

In this section, we review the computation of the meson spectrum in the
D4/D8-model.

The bosonic part of the action for a D-brane probe is
\begin{equation}
  S= S_{D B I}+S_{CS}+S_f,
\end{equation}
where the Abelian Dirac-Born-Infeld action is
\begin{equation}
	\label{eq:dbi}
  S_{DBI}= T \int \mathrm{d}^{p+1}\!\xi ~ e^{-\phi}
  \sqrt{-\det(g_{\alpha\beta}+\mathcal{F}_{\alpha\beta})},
\end{equation}
where 
$g_{\alpha\beta}=\partial_\alpha X^M\partial_\beta X^N G_{MN}$ 
is the induced metric,
$\mathcal{F}_{\alpha\beta}=2\pi\alpha'(F_{\alpha\beta}
+\mathcal{B}_{\alpha\beta})$, 
and 
$\mathcal{B}_{\alpha\beta}=\partial_\alpha X^M\partial_\beta X^N B_{MN}$ 
is the induced B-field and 
$F_{\alpha\beta}$ is the world-volume gauge field.
The constant $T$ is related to the string length $l_s$ by 
$T=((2 \pi)^8 l_s^9)^{-1}$.

The Chern--Simons part $S_{CS}$ is important \eg to achieve the correct
chiral anomaly of QCD as well as the Wess--Zumino--Witten term in the
chiral Lagrangian, but will not be relevant for us.
Our focus is on the fermionic part, $S_f$, which we will return to
after a very brief summary of meson spectroscopy.

Recall that the gauge sector of the field theory is associated with
open strings ending on the D4-branes, giving rise to a
$U(N_c)$ colour gauge symmetry, where $N_c$ is the number of
D4-branes. 
Introducing a probe brane means
introducing a flavour sector associated with strings stretched between
the D4 and the D8 or \anti{D8} branes, giving rise to a $U(N_f)$
global flavour symmetry, where $N_f$ is the number of probe branes.

The basic idea of the holographic description of mesons in this setup
is that they can be identified with fluctuations of the probe
D-branes.
This works straightforwardly for scalar and vector mesons, which are
associated with fluctuations of the probe D-brane collective
coordinates and world-volume gauge field. 
For higher spin mesons,
alternative approaches have to be applied. 
In general, the situation is
complicated, but ultra-high spin mesons have been studied in
refs.~\cite{Kruczenski:2004me,Peeters:2005fq} using a semi-classical
macroscopic spinning string description.

Returning to the low-spin case, in order to study the fluctuations it
is necessary to fix the coordinate invariance. 
This is done with a static gauge, which in our case is
\begin{equation}
  \xi^\alpha=X^\alpha, \quad \text{for } \alpha\neq 5,
  \qquad
  X^5(\xi) \equiv y(\xi),
\end{equation}
where the $\xi^\alpha$ are the nine-dimensional world-volume coordinates
and $X^M$ are ten-dimensional space-time coordinates.

The induced metric on the D8-brane in this case is
\begin{equation}
  ds^2_{D8} = 
  \left(\frac{U}{R}\right)^{\frac{3}{2}}\eta_{\mu\nu} dx^\mu dx^\nu
  +\frac{4}{9}\left(\frac{R}{U}\right)^\frac{3}{2}
  \frac{U_{kk}}{U} dz^2
  +R^{\frac{3}{2}}U^{\frac{1}{2}}d\Omega_4^2,
\end{equation}
where $U$ should now be viewed as a function of $z$ given by
\begin{equation}
  U(z)=U_{kk}\left(1+\frac{z^2}{U_{kk}^2}\right)^{1/3}.
\end{equation}
We will henceforth work with dimensionless quantities 
\bea
  w\equiv\frac{z}{U_{kk}}, \qquad V(w)\equiv(1+w^2)^\frac{1}{3}, 
  \qquad 
  \alpha\equiv\left(\frac{U_{kk}}{R}\right)^{\frac{3}{4}}=\left(\frac{2}{3}M_{kk}R\right)^{\frac{3}{2}}, 
\eea
such that the induced metric $g_{\alpha\beta}$ reads
\begin{equation}
   ds_{D8}^2 = g_{\alpha\beta}dx^\alpha dx^\beta 
             = \alpha^2V^\frac{3}{2} \eta_{\mu\nu} dx^\mu dx^\nu
             + \frac{\alpha^2}{M_{kk}^2} V^{-\frac{5}{2}}  dw^2
             + \frac{9}{4} \frac{\alpha^{2}}{M_{kk}^2} V^{\frac{1}{2}} d\Omega_4^2.
\end{equation}

The D8-brane degrees of freedom that remain after this gauge fixing is 
the scalar field $y$ (fluctuating around the value $y=0$) and the
world-volume gauge fields $A_\alpha$.
The nine-dimensional world-volume naturally splits up in a four-dimensional
part $x^\mu$, an extra dimension $w$ and an $S^4$ 
part. 
Considering only singlets under the $SO(5)$ of $S^4$, the gauge
field degrees of freedom are $A_\mu(x^\mu,w)$ and $A_w(x^\mu,w)$.

The Abelian Dirac--Born--Infeld action can be expanded up to second
order in the field strength as  
\bea
    S_{DBI} &=& T\int \mathrm{d}^9\xi \mathrm{e}^{-\phi}
                \sqrt{-\det(\mathrm{g_{\alpha\beta}}+{\mathcal F}_{\alpha\beta})}\nn\\ 
            &=& \frac{\hat{T} \alpha }{(2 \pi \alpha')^2} \int \mathrm{d}^4x\mathrm{d}w~ 
                V^2~ \left( 1 + (2 \pi \alpha')^2\frac{1}{4 \alpha^4} V^{-3} \eta^{\mu\nu}\eta^{\rho\sigma}
	        F_{\mu\nu} F_{\rho\sigma}\right. \nn\\
 	     && \qquad\qquad \left.+ (2 \pi \alpha')^2\frac{M_{kk}^2}{2 \alpha^4} 
	        V \eta^{\mu\nu} F_{\mu w} F_{\nu w} \right)+ {\mathcal O}(F^3), 
\eea
where
\begin{equation}
  \label{eq:hatT}
  \hat{T} \equiv 
  \left(\frac{3}{2}\right)^4\frac{TV_4(2 \pi \alpha')^2}{g_s}\frac{\alpha^7}{M_{kk}^5}.
\end{equation}
is a dimensionless constant.

We are interested in finding a four-dimensional effective action for
the gauge field fluctuations with a canonical normalisation.
To achieve this we expand the gauge fields and $y$ in terms of
complete sets  $\{p_n(w)\}$, $\{q_n(w)\}$ and $\{\rho_n(w)\}$,
\begin{align}
\label{eq:b}
  A_\mu(x^\mu,w) &= \sum_n \mathcal{V}_\mu^{(n)}p_n(w),
  \\
  A_w  (x^\mu,w) &= \sum_n \varphi^{(n)}q_n(w),
  \\
  y    (x^\mu,w) &= \sum_n \mathcal{U}^{(n)}(x^\mu)\rho_n(w).
\end{align}

In the following we will focus on the vector fields and assume
$N_f=1$.
The expansions of the field strengths then take the form 
\begin{align}
  F_{\mu\nu}(x,w) &  
  = \sum_n \left( \partial_\mu {\cal V}^{(n)}_\nu(x)
  -\partial_\nu {\cal V}^{(n)}_\mu(x)\right)p_n(w)
  \nonumber
  \\ &
  \equiv \sum_n F_{\mu\nu}^{(n)}(x)p_n(w)
  \\
  F_{\mu  w}(x,w) &
  = \sum_n \partial_\mu \varphi^{(n)}(x) q_n(w)-{\cal
    V}^{(n)}_\mu(x)p'_n(w).
\end{align}
In order to have canonically normalised kinetic terms in the effective
four-dimen\-sional action, we impose the normalisation conditions
\begin{align}
  \label{eq:normalisationF}
    \frac{\hat{T}}{\alpha^3} \int \mathrm{d}w~ V^{-1}(w)p_n(w) p_m(w)
  =& \delta_{nm},
  \\
   \frac{\hat{T} M_{kk}^2}{\alpha^3} \int \mathrm{d}w~ V^{3}(w) q_n(w) q_m(w)
  =& \delta_{nm},
\end{align}
and choose the $p_n(w)$ to be eigenfunctions of the equation  
\bea
   \label{eq:eig}
   -V\partial_w(V^3\partial_w p_n(w))=\lambda_n p_n(w).
\eea
This equation together with eq.~\eqref{eq:normalisationF}
implies
\bea
  \label{eq:normdiff}
   \frac{\hat{T}}{\alpha^3} 
  \int \mathrm{d}w~ V^3(w) p'_n(w) p'_m(w)=\lambda_n\delta_{nm}.
\eea
Comparing eqs.~\eqref{eq:normalisationF} and \eqref{eq:normdiff}, 
we notice that we can choose the eigenfunctions $q_n$ to be proportional to $p'_n$:
\bea
   q_n=\left(M_{kk}^2 \lambda_n \right)^{-1/2}p'_n.
\eea
The fields $\varphi^{(n)}(x)$ can be absorbed into the vector meson
field by the gauge transformation 
\bea
   {\cal V}^{(n)}_\mu(x) \to {\cal V}^{(n)}_\mu(x)
   +\left(\lambda_n M_{kk}^2 \right)^{-1/2} 
   \partial_\mu \varphi^{(n)}(x), 
\eea
such that the action for the vector meson field 
${\cal V}^{(n)}_\mu(x)$ takes the canonical form  
\bea
  S=\int \mathrm{d}^4x~ \left[ \sum_{n\geq1} \left( 
  \frac{1}{4} F_{\mu\nu}^{(n)}F^{\mu\nu(n)}
  +\frac{1}{2} \lambda_n M_{kk}^2 
  {\cal V}_{\mu}^{(n)} {\cal V}^{\mu (n)}
  \right)\right]
\eea
for the massive vector fields ${\cal V}$.
The field strength mode expansion is given by
\bea
  F_{\mu  w}(x,w) &
  = -\sum_n {\cal V}^{(n)}_\mu(x)p'_n(w)\equiv\sum_n F_{\mu  w}^{(n)}(x)p'_n(w).
\eea
Note that the eigenfunctions $p_m$ are chosen to be either odd or even
functions 
\bea
\label{eq:pparity}
    p_m(-w)=(-)^{m+1}p_m(w),
\eea
such that the odd numbered modes are vectors whilst the even ones are
axial vectors. 

In addition to the above massive modes, there are non-normalisable
solutions to the eq.~\eqref{eq:eig} with eigenvalue zero: 
\bea
   p_0(w)=C \arctan(w).
\eea
This reflects the fact that there exist a function $q_0=C V^{-3}$,
which is orthogonal to all other $q_n$ 
\bea
  \int \mathrm{d} w~ V^3 q_0 q_n \propto \int \mathrm{d} w~  p'_n=0. 
\eea
The normalisation constant is determined easily to be
\bea
  \label{eq:p0norm}
  C= \left( \frac{\hat{T} M_{kk}^2}{\alpha^3} 
  \pi \right)^{-1/2}.
\eea
Taking the zero mode into account, the effective action becomes
\bea
  S=\int \mathrm{d}^4x~ \left[ 
    \frac{1}{2} \partial_\mu \varphi^{(0)}\partial^\mu \varphi^{(0)} 
    + \sum_{n\geq1} \left( \frac{1}{4} F_{\mu\nu}^{(n)}F^{\mu\nu (n)}+\frac{1}{2} \lambda_n M_{kk}^2 {\cal V}_{\mu}^{(n)} {\cal V}^{\mu (n)}\right)\right],
\eea
where the massless scalar field $\varphi^{(0)}$ is interpreted as the pion.

The masses of the vector mesons were found in
ref.~\cite{Sakai:2004cn} using a numerical shooting technique. 
These masses can be matched with the observed values amazingly
well. 
It is surprising because we have no really good understanding of
why this particular D-brane setup should be a good approximation of
the holographic dual of QCD.

So far, we have tacitly assumed that the gauge potential vanishes as
$w\to \pm \infty$.  
The field strength should vanish as $w\to \pm \infty$ for the
effective action to be normalisable. 
Then it is always possible to choose a gauge where the potential
vanishes for large $|w|$. 
In section~\ref{sec:mes} we shall study couplings to an external gauge field, and for that it is useful to employ the gauge where $A_w=0$.
At first look the pion field $\varphi^{(0)}$ seems to be gauged away
in this gauge but this is impossible.  
There exists no gauge, which sets $A_w=0$ and lets $A_\mu$ vanish for
large $|w|$ at the same time. 

The $A_w=0$ constraint can be accomplished by the gauge transformation
\bea
  \label{eq:gaugetrans}
   A_\alpha \to A_\alpha -\partial_\alpha \Lambda
\eea
with the gauge function $\Lambda$ given by
\bea
  \Lambda(x,w)= \varphi^{(0)}(x) p_0(w) +\sum_{n=1}^\infty \varphi^{(n)}(x) \left( M_{kk}^2 \lambda_n\right)^{-1/2} p_n(w).
\eea
After the gauge transformation eq.~(\ref{eq:gaugetrans}), the gauge potential reads
\bea
   A_\mu(x,w)\! &=&\! -\partial_\mu \varphi^{(0)}(x) p_0(w)+ \sum_{n\geq1} \left({\mathcal V}_\mu^{(n)}(x) - \frac{1}{M_{kk}\sqrt{\lambda_n}} \partial_\mu \varphi^{(n)}(x) \right)p_{n}(w), \nn
  \\
  A_w(x,w)   \! &=&\! 0.
\eea
The contribution from the massive scalar modes $ \varphi^{(n)}$ can again be absorbed into the massive vectors ${\mathcal V}_\mu^{(n)}$ by a field redefinition. 
The asymptotic values of the gauge potential for large values of $|w|$ are
\bea
   A_\mu(x,-\infty) 
   =  C \frac{\pi}{2} \partial_\mu \varphi^{(0)}(x), \\
   A_\mu(x, \infty)
   = -C \frac{\pi}{2} \partial_\mu \varphi^{(0)}(x).
\eea

Taking an external gauge field into account, the gauge potential in
$A_w=0$ gauge may be written as 
\bea
   A_\mu(x,w) &=& \mathcal{\hat V}_\mu +\mathcal{\hat A}_\mu p_0(w)+ \sum_{n\geq1} {v}_\mu^{(n)}p_{2n-1}(w)+ \sum_{n\geq1} {a}_\mu^{(n)}p_{2n}(w),
  \\
  A_w(x,w) &=& 0,
\eea
where the fields $v_\mu^{(n)}$ and $a_\mu^{(n)}$ are the massive
vector and axial vector mesons, respectively. 
The zero modes are given by 
\bea
  \mathcal{\hat A}_\mu&=&\frac{1}{2}\left(A_{L\mu}-A_{R\mu}\right)+ \frac{i}{f_\pi}\partial_\mu \pi,\\
  \mathcal{\hat V}_\mu&=&\frac{1}{2}\left(A_{L\mu}+A_{R\mu}\right),
\eea
where $A_{L\mu},\ A_{R\mu}$ are the asymptotic values of the external
gauge field for $w\to \pm \infty$ and the pion field $\varphi^{(0)}
\equiv i \pi$ with its decay constant $f_\pi \equiv 2/(\pi~C)$. 
In particular, the electromagnetic photon field can be extracted by
setting $A_{L\mu}=A_{R\mu}=eQA_\mu^{EM}$ for the electric charge $Q$. 
The field strengths read
\bea
 \label{eq:fmn}
  F_{\mu\nu}(x,w)&=& \partial_\mu \hat{\cal  V}_\nu(x)-\partial_\nu \hat{\cal V}_\mu(x) 
                   + \left(\partial_\mu  \hat{\cal A}_\nu(x)-\partial_\nu \hat{\cal A}_\mu(x) \right) p_0(w)\nn\\
                   &&+ \sum_{n\geq 1} \left(\partial_\mu v^{(n)}_\nu(x)-\partial_\nu v^{(n)}_\mu(x)\right)p_{2n-1}(w)\nn\\
                   &&+ \sum_{n\geq 1} \left(\partial_\mu a^{(n)}_\nu(x)-\partial_\nu a^{(n)}_\mu(x)\right)p_{2n}(w)\nn\\
            &\equiv& \sum_n F_{\mu\nu}^{(n)}(x)p_n(w),\\
 \label{eq:fmw}
  F_{\mu  w}(x,w)&=&{\cal \hat A}_\mu(x) p_0'(w) + \sum_{n\geq 1} v^{(n)}_\mu(x)p_{2n-1}'(w)
                   + \sum_{n\geq 1} a^{(n)}_\mu(x)p_{2n}'(w)\nn
    \\        &\equiv&\sum_n  F_{\mu  w}^{(n)}(x) p'_n(w),
\eea
where we indicated the split into vector and axial components.
Equipped with these expansions we can aim for effective couplings of
mesinos to gauge fields. 

\section{Mesino spectroscopy}

As promised in the introduction, we shall now turn to the fermionic
sector of the probe D-branes and compute the masses of these
fluctuations.
A similar calculation has been partly done in the appendix of
ref.~\cite{Sakai:2004cn}. 
The same ideas have been applied also in ref.~\cite{Kirsch:2006he} for
computing the masses of the meson superpartners in the D3/D7 setup. In
that model, $\mathcal{N}=2$ supersymmetry is preserved and the masses
can be calculated analytically. 
As expected, there is a match between fermion and boson masses.

The fermionic part of the D8-brane action is given by~
\cite{Marolf:2003ye,Marolf:2003vf,Martucci:2005rb}
\bea
   \label{eq:fermionicaction}
   S^{(f)}=i T\int \rm{d}^9\xi \sqrt{-(g+{\cal F})} \rm{e}^{-\phi} 
            \bar \Psi \frac{1}{2}(1-\Gamma_{D8})(\Gamma^\alpha D_\alpha -\Delta+L_{D8})\Psi,
\eea 
where
\begin{align}
  \Gamma_{D8}&=\frac{\sqrt{-g}}{\sqrt{-(g+{\cal F})}}
  \Gamma_{D8}^{(0)} \Gamma^{11}\sum_q \frac{(-\Gamma^{11})^q}{q!2^q} 
  \Gamma^{\alpha_1\dots\alpha_{2q}} {\cal F}_{\alpha_1\alpha_2}
  \dots {\cal F}_{\alpha_{2q-1}\alpha_{2q}},
  \\
  L_{D8}&= \frac{-\sqrt{-g}}{\sqrt{-(g+{\cal F}})}\Gamma_{D8}^{(0)}
  \sum_{q \geq 1} 
  \frac{(-\Gamma^{11})^{q-1}}{(q-1)! 2^{q-1}}
  \Gamma^{\alpha_1\dots\alpha_{2q-1}} {\cal F}_{\alpha_1\alpha_2}\dots
  {\cal F}_{\alpha_{2q-1}}^{\phantom{\alpha_{2q-1}}\beta} D_\beta,
  \\
  \Gamma_{D8}^{(0)}&=
  \frac{\epsilon^{\alpha_1\dots\alpha_9}}{9!\sqrt{-g}}
  \Gamma_{\alpha_1\dots\alpha_9} 
  =-\Gamma^{11}\Gamma^{\underline5},
  \\
  \Delta &= \frac{1}{2}\Gamma^M\partial_M\phi
  -\frac{1}{8}\frac{1}{4!}e^\phi F_{N P Q R}\Gamma^{N P Q R},   
  \\
  D_M &= \nabla_M 
  -\frac{1}{8}\frac{1}{4!}e^\phi F_{N P Q R}\Gamma^{N P Q R}\Gamma_M,   
  \\
  \nabla_M &=\partial_M+\frac{1}{4} \Omega_M^{\underline{NP}}
  \Gamma_{\underline{NP}}.
\end{align}
The $\Omega$'s are the spin connections, 
$\nabla_M=\partial_M+\frac{1}{4}\Gamma^{AB}\Omega_{ABM}$ 
is the usual covariant derivative and 
$\Gamma^{N P Q R}=\Gamma^{[N}\Gamma^P\Gamma^Q\Gamma^{R]}$.
Latin indices $M=0,\dots,9$ refer to space-time and Greek indices
$\alpha=0,\dots,4,6,\dots 9$ refer to the D8-brane world-volume. 
Quantities are pulled back onto the D8-brane in the usual way, 
\eg
$\nabla_\alpha = \partial_\alpha X^M\nabla_M$.
Transformations between curved indices $\alpha,M$ and flat (local Lorentz frame) indices
$\underline{\alpha},\underline{M}$ are done with the vielbeins, so we
have \eg
$\Gamma^\alpha = g^{\alpha\beta}\partial_\beta X^M E_M^{~~\underline{M}} 
\Gamma_{\underline{M}}$.

The field $\Psi$ is a 32-component Majorana spinor in ten dimensions and
\begin{equation}
  P_- \equiv\frac{1}{2}(1-\Gamma_{D8})
\end{equation}
is a projection operator necessary for kappa invariance.
The $\Gamma_{D8}$ matrix is in the case of vanishing gauge field (${\mathcal F}_{\alpha \beta}=0$)
\begin{equation}
  \Gamma_{D8} \equiv \frac{1}{9!}\frac{1}{\sqrt{-g}}
  \epsilon^{\alpha_1\cdots\alpha_9} 
  \Gamma_{\alpha_1\cdots\alpha_9} \Gamma^{11}
  =\Gamma_{\underline{5}}.
\end{equation}

We are dealing with Majorana spinors in ten dimensions.
These spinors obey a reality condition.
For the product of two Majorana spinors with an arbitrary number of
$\Gamma$ matrices, we have 
\begin{equation}
  \bar\chi \Gamma^{M_1\dots M_n} \Psi 
  = (-)^{n(n+1)/2} \bar\Psi \Gamma^{M_1\dots M_n} \chi
\end{equation}
so that these expression vanish if the spinors 
are the same and $n=1,2,5,6,9,10$. 
Moreover, choosing as the $\kappa$ symmetry fixing condition
the chirality condition 
$\Gamma^{11} \Psi = \Psi$, 
then the
only non-vanishing products are 
\bea
  \bar\Psi \Gamma^{M_1M_2M_3} \Psi,
  \qquad
  \bar\Psi \Gamma^{M_1 \dots M_7} \Psi,
\eea
since the chirality matrix $\Gamma^{11}$ transforms products of four and eight 
$\Gamma$ matrices into six and two, which vanish.

As mentioned before, setting $y=0$ is equivalent to setting the quark
mass to zero. 
This is an assumption we are making throughout this paper.
Since we are now not interested in the fluctuations of the bosonic fields, 
we set $\mathcal{F}_{\alpha\beta}=0$ in the calculation of the mesino spectrum 
and also set $\partial y=0$.
We will include the gauge field contributions when considering mesino interactions 
with vector mesons in the next section.
We find that the contributions from the flux $F_4$ cancel in the action
\eqref{eq:fermionicaction}. 
The measure and dilaton factors together give
\begin{equation}
  T \sqrt{-g}e^{-\phi}
  =\hat{T}(V_4(2 \pi \alpha')^2)^{-1} \alpha V^2
  \end{equation}
With the definitions given above and some rather straightforward
manipulations, we find
\begin{equation}
  S = i\hat{T}(V_4 (2 \pi \alpha')^2)^{-1}
  \int \mathrm{d}^4\!x \mathrm{d}w d\Omega_4
  V^2 \mathcal{L}, 
\end{equation}
where
\begin{equation}
\label{eq:a}
\begin{split}
  \mathcal{L} = \bar{\Psi}P_{-}\Bigl[ &
  \frac{2}{3} M_{kk} V^{-\frac{1}{4}} 
  \Gamma^{\underline{m}} \nabla_m^{S^4}
  +V^{-\frac{3}{4}}  \Gamma^{\underline{\mu}}\partial_{\mu}
  \\ &
  + M_{kk} V^{\frac{5}{4}}  \Gamma_{\underline{4}}\partial_w
  +\frac{13}{12} M_{kk} w V^{-\frac{7}{4}} 
  \Gamma_{\underline{4}}
  \Bigr]\Psi,
\end{split}
\end{equation}
where $m=6,7,8,9$ and $\Gamma^{\underline{m}} \nabla_m^{S^4}$ is the
Dirac operator on a unit four-sphere. 

We rescale the spinor 
\bea
  \label{eq:spinorrescsale}
  \Psi\to \tilde \Psi=V^{-13/8}\Psi.
\eea
This rescaling removes the last term in Eq.(\ref{eq:a}).
Moreover, it leads to an action where the term with the $w$ derivative
has no weighting factor $V$ 
\begin{equation}
  S =\frac{i\hat{T}}{V_4 (2 \pi \alpha')^2}
    \int \mathrm{d}^4\!x \mathrm{d}w d\Omega_4 ~\bar{\Psi}P_{-}\left[ 
   \frac{2}{3} M_{kk}V^{-\frac{3}{2}} \Gamma^{\underline{m}} \nabla_m^{S^4}
  +  V^{-2}\Gamma^{\underline{\mu}}\partial_{\mu}
  +M_{kk} \Gamma_{\underline{4}}\partial_w
  \right]\Psi.
\end{equation}

\subsection{Spinor decomposition}

We want to decompose our spinor $\Psi$ into an $S^4$ part $\chi$, a $3+1$-dimensional part $\psi$ and 
a remaining two-dimensional part $u$, \ie
\begin{equation}
\label{eq:c}
  \Psi = u \otimes \psi(x,z) \otimes \chi(S^4).
\end{equation}
A convenient choice for the decomposition of the 10-dimensional gamma matrices is 
\begin{subequations}
\begin{align}
  \Gamma^{\underline{\mu}} &
  =\sigma_x\otimes\gamma^{\underline{\mu}}\otimes\unitmatrix,
  \qquad \mu=0,1,2,3;
  \\
  \Gamma^{\underline{4}} &
  =\sigma_x\otimes\gamma\otimes\unitmatrix,
  \\
  \Gamma^{\underline{5}} &
  =\sigma_y\otimes\unitmatrix \otimes \tilde \gamma,
  \\
  \Gamma^{\underline{m}} &
  =\sigma_y\otimes\unitmatrix\otimes\tilde{\gamma}^{\underline{m}},
  \qquad m=6,7,8,9;
\end{align}
\end{subequations}
where 
$\gamma = i\gamma^{\underline{0}}\gamma^{\underline{1}}
\gamma^{\underline{2}}\gamma^{\underline{3}}$
is the chiral matrix in $3+1$ dimensions and
$\tilde{\gamma} = 
\tilde{\gamma}^{\underline{6}}
\tilde{\gamma}^{\underline{7}}
\tilde{\gamma}^{\underline{8}}
\tilde{\gamma}^{\underline{9}}$
is the chiral matrix on the tangent space of $S^4$.
In this decomposition, the ten-dimensional chirality matrix takes a 
particularly simple form: $\Gamma^{11}=\sigma_z \otimes\unitmatrix \otimes \unitmatrix$.
If we further decompose the $u$ into $\sigma_z$ eigenstates $u_\pm$,
we get
\begin{equation}
  \sigma_z u_\pm = u_\pm, \qquad
  \sigma_x u_\pm = u_\mp, \qquad
  \sigma_y u_\pm = \pm i u_\mp.
\end{equation}
The kappa symmetry fixing condition $\Gamma^{11}\Psi=\Psi$ is then
equivalent to 
$u=u_+ \equiv \ket{\uparrow}$.

The general solution for the Dirac equation on the four sphere is
well-known \cite{Camporesi:1995fb},
\begin{equation}
  \fslash{\nabla}_{\!\!\!S^4}~\chi^{\pm ls}=i \lambda^\pm_l~\chi^{\pm ls};
  \qquad
  \lambda^\pm_l=\pm (2+l);\quad l=0,1,\dots,
\end{equation}
where the quantum number $s=0,1,\dots, d_l$ and 
$d_l=4(3+l)!/(3!~l!)$ is the
degeneracy of the eigenvalue $\lambda^\pm_l$. 
With the above splitting and
$\bar{\Psi}=\bra{\uparrow}\sigma_x\otimes\bar{\psi}\otimes\chi^\dagger$, 
the Lagrangian becomes
\begin{equation}
  \mathcal{L} \sim
  \braket{\uparrow}{\uparrow}~\chi^\dagger \chi~
  \bar{\psi}\bigl[ \dots \bigr] \psi.
\end{equation}

Normalising the fields according to
\begin{equation}
  \braket{\uparrow}{\uparrow}\int
  d\Omega_4~(\chi^{ls})^\dagger \chi^{l's'} 
  = (2 \pi \alpha')^2 V_4\delta^{l,l'}\delta^{s,s'},
\end{equation}
we arrive at the action
\bea
  \label{eq:reducedaction}
  S = i\hat{T}\int \mathrm{d}^4\!x \mathrm{d}w~ \bar{\psi} \left[     
    - \frac{2}{3} M_{kk}\lambda V^{-\frac{3}{2}}
    + V^{-2} \gamma^{\underline{\mu}}\partial_\mu
    + M_{kk} \gamma \partial_w 
    \right] \psi.
\eea

\subsection{Four-dimensional action in canonical form}

In order to read off the mass of the four-dimensional fluctuations, we
need to rewrite the action \eqref{eq:reducedaction} in canonical
form. 
We do this by working with two-spinors $\psi_\pm$, 
\begin{equation}
\label{eq:weylspinor}
  \psi=\left(\begin{matrix}\psi_+\\ \psi_- \end{matrix} \right),
\end{equation}
and choose the Weyl basis where
\begin{equation}
\label{eq:weylbasis}
  \gamma^\mu=i \left(\begin{matrix} 
    0 & \sigma^\mu \\ 
    \bar\sigma^\mu & 0 
  \end{matrix} \right),
 \quad
  \gamma=\left(\begin{matrix}
    \unitmatrix & 0\\
    0 & -\unitmatrix \end{matrix} \right),
\end{equation}
where $\sigma^\mu=(\unitmatrix, -\sigma^i)$ and 
$\bar\sigma^\mu=(\unitmatrix, \sigma^i)$.
Then we expand the fermion fields in terms of complete
sets $\{f_+^n(w)\}$ and $\{f_-^n(w)\}$,
\begin{equation}
  \label{eq:expansion}
  \psi_+(x^\mu,w) = \sum_n \psi_+^{(n)}(x^\mu) f_+^n(w),
  \quad
  \psi_-(x^\mu,w) = \sum_n \psi_-^{(n)}(x^\mu) f_-^n(w),
\end{equation}
where the functions $f_+^n$ and $f_-^n$ are real eigenfunctions 
of the coupled first-order differential equations
\begin{equation}
\begin{split}
  \label{eq:diffeqs}
  -\frac{2\lambda}{3} V^{-\frac{3}{2}} f_+^n(w)
  +\partial_w f_+^n(w)
  = & \tilde{M_n} V^{-2} f_-^n(w),
  \\
  -\frac{2\lambda}{3} V^{-\frac{3}{2}} f_-^n(w)
  -\partial_w f_-^n(w)
  = &  \tilde{M_n} V^{-2} f_+^n(w),
\end{split}
\end{equation}
where $\tilde{M}_n \equiv \frac{M_n}{M_{kk}}$.
With the normalisations
\begin{equation}
  \label{eq:normalisation}
   \hat T \int \mathrm{d}w~ V^{-2}(w)f_\pm^{n}(w) f_\pm^m(w)=\delta^{nm},
\end{equation}
the action takes the canonical form 
\begin{equation}
  \!S\! =\! -\!\int\! \mathrm{d}^4 x\! \sum_n\! \left\{
  \psi_-^{(n)\dagger} i\sigma^\mu\partial_\mu \psi^{(n)}_-
  +\psi_+^{(n)\dagger}  i\bar{\sigma}^\mu\partial_\mu\psi_+^{(n)}
  +M_n\left[ \psi_-^{(n)\dagger} \psi_+^{(n)}
  +\psi_+^{(n)\dagger} \psi_-^{(n)} \right]
  \right\},
\end{equation}
which is the action, written in the Weyl basis, for a set of Dirac
spinors  
\begin{equation}
  \label{eq:mesinospinor}
  \tilde\psi^{(n)}=\begin{pmatrix} \psi_+^{(n)} \\ \psi_-^{(n)} \end{pmatrix}
\end{equation}
with masses $M_n$.
With these Dirac spinors, the action takes the familiar form
\bea
S = i\int \mathrm{d}^4 x \sum_n \left[
  \bar{\tilde \psi}^{(n)} \gamma^\mu\partial_\mu \tilde \psi^{(n)}
  +M_n \bar{\tilde \psi}^{(n)} \tilde \psi^{(n)}
   \right].
\eea
In order to compute the masses $M_n$ we have to solve the eigenvalue
problem \eqref{eq:diffeqs}.
We will do this numerically in the next section.

The coupled first order differential equations \eqref{eq:diffeqs}
can be rewritten as two decoupled second order differential equations
in Sturm-Liouville 
form, 
\begin{equation}
  -\partial_w\bigl[ V^2\partial_w f^n_\pm(w)\bigr]
  +q_\pm(w) f^n_\pm(w) 
  = \tilde{M}_n^2 V^{-2} f^n_\pm(w),
\end{equation}
where
\begin{equation}
  q_\pm(w) = \frac{4 \lambda^2}{9} V^{-1} \pm \frac{2\lambda}{9}w V^{-5/2}
           = \frac{w^2}{36} V^{-4} \left[
	   \left( 4 \lambda \sqrt{1+\frac{1}{w^2}} \pm 1 \right)^2 
	   - 1 \right] 
	   \geq 0.
\end{equation}
We therefore know that the eigenvalues $\tilde{M}_n$ are real and
well-ordered and that the eigenfunctions are real and unique.
At this point we note that $\lambda\to -\lambda$ is equivalent to
swapping $f^n_\pm \to f^n_\mp$, so we can therefore assume $\lambda>0$
without loss of generality.

Inspecting the equations, we note that $f_-^n(-w)$ satisfies the same
equation as $f_+^n(w)$. The uniqueness of the eigenfunctions then
implies that they are the same up to a sign, \ie we have
\begin{equation}
  \label{eq:plusminussymmetry}
  f_\mp^n(-w) = c^{(n)} f_\pm^n(w), \qquad c^{(n)}=\pm 1.
\end{equation}

For the special case $M=0$, the equations \eqref{eq:diffeqs} can be
solved analytically, giving
\begin{equation}
  f_\pm = C_\pm (w+\sqrt{w^2+1})^{\pm \frac{2\lambda}{3}} .
\end{equation}
However, this is a non-normalisable solution. 

\subsection{Numerical results}

The asymptotic behaviour of the eigenfunctions is found by
investigating the differential equations \eqref{eq:diffeqs} for
$w\to\infty$ and yields 
\begin{equation}
\begin{split}
  \label{eq:asymptotic}
  f_+ &\sim
	\left( A~ w^{-\frac{2\lambda}{3}-\frac{1}{3}} 
  - B \frac{4\lambda-1}{3\tilde{M}}~ w^{\frac{2\lambda}{3}}
  \right),
  \\
  f_- &\sim 
	\left( - A \frac{4\lambda+1}{3\tilde{M}} ~ w^{-\frac{2\lambda}{3}} 
  + B~ w^{\frac{2\lambda}{3}-\frac{1}{3}}
  \right),
\end{split}
\end{equation}
where $A$ and $B$ are undetermined constants.
As we have already noted, it is sufficient to consider the $\lambda=(l+2)\ge 2$ case. 
Then we have
\begin{equation}
  f_+ f_+ \sim B^2 w^\frac{4\lambda}{3},\qquad f_- f_- \sim B^2 w^{\frac{2}{3}(2\lambda-1)}
\end{equation}
This expression is not normalisable according to equation
\eqref{eq:normalisation} unless we require $B=0$. It is this
regularity condition, which produces a discrete mass spectrum.

Let us now turn to the initial conditions.
Evaluating the equations \eqref{eq:diffeqs} at $w=0$ gives
\begin{equation}
  \label{eq:equationsatzero}
   \pm {f^n_\pm}'(0) - \frac{2\lambda}{3} f_\pm^n(0) = \tilde{M}_n f_\mp^n(0).
\end{equation}
From equation \eqref{eq:plusminussymmetry} we get
\begin{equation}
  \label{eq:symmetryatzero}
  f_-(0)= c f_+(0),
  \quad
  f_-'(0)= -c f_+'(0).
\end{equation}
Equations \eqref{eq:equationsatzero} and \eqref{eq:symmetryatzero}
together give
\begin{equation}
\begin{array}{ll}
  f_+(0)= f_0; &  f'_+(0)=(\frac{2\lambda}{3} -c \tilde{M})f_0;
  \\
  f_-(0)= c f_0; & f'_-(0)= -c (\frac{2\lambda}{3} -c \tilde{M})f_0,
\end{array}
\end{equation}
so that the only choice we have in initial
conditions is in $f_0$ and the sign of $c$.

The aim when solving the eigenvalue problem numerically is to identify
the mass eigenvalues $\tilde{M}_n$ corresponding to regular
eigenfunctions.
In doing this we apply a shooting method.
The actual value of $f_0$ is not important and can always be set to
$f_0=1$. 
The recipe is then to
choose either $c=+1$ or $c=-1$ and guess a value of $\tilde{M}$.
Then we integrate the equations numerically up to very large $w$ and compute the coefficient $B$ in the
asymptotic behaviour \eqref{eq:asymptotic} according to 
\begin{equation}
  B = \lim_{w\to\infty} 
  w^{-\frac{2\lambda}{3}+\frac{1}{3}} f_-(w).
\end{equation}
\begin{figure}
  \begin{center}
    \includegraphics{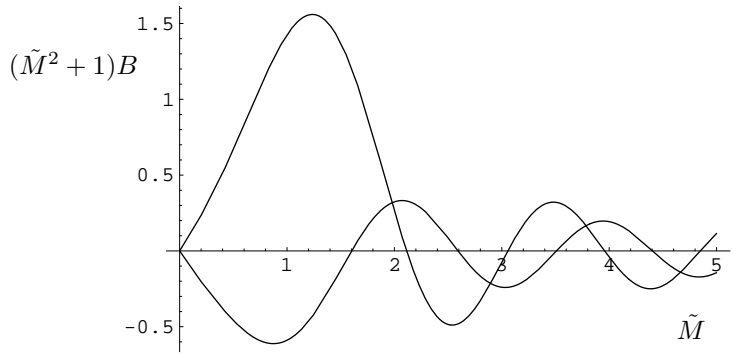}
    \caption{The coefficient $B$ plotted as a function of $\tilde{M}$
      for  $\lambda=2$ and $c=\pm 1$. The intersection points $B=0$ give
      the mass eigenvalues.
      \label{fig:eigenvalues}}
  \end{center}
\end{figure}
Different $\tilde{M}$ give different values for $B$, and as we have
argued above, only those giving $B=0$ are regular mass eigenstates,
see figure~\ref{fig:eigenvalues}. 
In order to determine all the eigenvalues, we scan through a range of
$\tilde{M}$ for $c=\pm 1$ and for various $l \ge 0$.
It turns out that the lightest masses are found for $c=-1$. In
general, odd (even) $n$ corresponds to $c=-1$ ($c=+1$):
\begin{equation}
  \label{eq:fparity}
  f_\mp^n(-w)=(-)^{n} f_\pm^n(w).
\end{equation}
Numerical values for $\tilde{M}$ are presented in Table~\ref{table}.

We can determine the parity of the eigenstates in the same way as in
ref.~\cite{Sakai:2004cn}.
Consider a transformation $L: (t,\vec{x},w)\to (t,-\vec{x},-w)$, which is
a proper Lorentz transformation in five dimensions.
The four-dimensional interpretation is that of a parity transformation. 
Acting on a spinor, the transformation is
\begin{equation}
  \psi(t,\vec{x},w) \to (-i\gamma^{\underline{0}})\psi(t,-\vec{x},-w),
\end{equation}
which, using the expansion \eqref{eq:expansion} gives
\begin{equation}
  \begin{pmatrix} 
    \psi_+ \\
    \psi_- 
  \end{pmatrix}
  \to \begin{pmatrix} 
    \sum_n\psi^{(n)}_-(t,-\vec{x}) f^n_-(-w) \\
    \sum_n\psi^{(n)}_+(t,-\vec{x}) f^n_+(-w) 
  \end{pmatrix}
  =  \begin{pmatrix} 
    \sum_n(-)^{n}\psi^{(n)}_-(t,-\vec{x}) f^n_+(w) \\
    \sum_n(-)^{n}\psi^{(n)}_+(t,-\vec{x}) f^n_-(w) 
  \end{pmatrix},
\end{equation}
where we have used the relation \eqref{eq:fparity}.
In terms of the mesino field
$\tilde{\psi}^{(n)}$ defined in equation~\eqref{eq:mesinospinor}, we see
that this has the effect
\begin{equation}
  \tilde{\psi}^{(n)}= 
  \begin{pmatrix} 
    \psi^{(n)}_+ \\
    \psi^{(n)}_- 
   \end{pmatrix}
   \to
   (-)^{n}\begin{pmatrix} 
    \psi^{(n)}_- \\
    \psi^{(n)}_+ 
   \end{pmatrix}
   =
   (-)^{n}(-i\gamma^{\underline{0}})
   \begin{pmatrix} 
    \psi^{(n)}_+ \\
    \psi^{(n)}_- 
   \end{pmatrix}.
\end{equation}
From this result we see that even-numbered modes correspond to mesinos with even parity, while odd-numbered ones have odd mesino parity.
The parity of the lowest mass eigenstates is indicated with a sign in
Table~\ref{table}.

For comparison, the massless scalar meson (the pion) has negative parity,
as does the lightest vector meson ($\rho$), whilst the lightest scalar
meson, $a_0(1450)$, has positive parity. See ref.~\cite{Sakai:2004cn}.

\begin{table}
\begin{center}
\begin{tabular}{l|llllll}
  & $n=1$ & $n=2$ & $n=3$ & $n=4$ & $n=5$ & $n=6$
  \\ \hline
  $l=0$ & $1.60^-$ & $2.11^+$ & $2.59^-$ & $3.05^+$ & $3.51^-$ & $3.96^+$
  \\
  $l=1$ & $2.27^-$ & $2.80^+$ & $3.29^-$ & $3.77^+$ & $4.24^-$ & $4.70^+$ 
  \\
  $l=2$ & $2.94^-$ & $3.48^+$ & $3.98^-$ & $4.47^+$ & $4.95^-$ & $5.41^+$
  \\
  $l=3$ & $3.61^-$ & $4.15^+$ & $4.67^-$ & $5.17^+$ & $5.65^-$ & $6.13^+$
\end{tabular}
\end{center}
\caption{\label{table}Mesino masses in units of $M_{kk}$ with parity
  indicated.}
\end{table}

It is interesting to compare the mass spectrum to that of the ordinary
mesons. The mass square ratio between the lightest massive mesino and
the lightest massive meson (the $\rho$ meson, see
ref.~\cite{Sakai:2004cn}) is 
\begin{equation}
  \frac{M^2}{m_\rho^2} = \frac{1.60^2 M_{kk}^2}{0.67 M_{kk}^2}
  = 3.8.
\end{equation}
In comparison, the mass square ratio of the second lightest and the
lightest meson is $2.4$ (from this D4/D8-brane model) / $2.51$
(experimentally). 
What we have found is that the mesino masses are comparable to the
meson masses.

\section{Mesino interactions}

We have found that the model predicts fermionic partners to the
mesons, which we have called mesinos. Their masses are of the same
order as meson masses, yet they have no counterpart in QCD or in
experiments. This by itself presents a problem for this model as a
holographic dual of QCD.
However, there is still the possibility that these mesinos do not
interact with other particles. If this is the case, we would be able
to view this fermionic sector as irrelevant since it does not affect the
meson sector. 
This section is devoted to the question of whether this is in fact the
case. 
Specifically, we will investigate the coupling of the mesinos to vector mesons
 as well as to an external gauge field.
  
As in previous sections, we shall consider only $SO(5)$ singlet states.
Thus we set $A_m=0,~m=6,7,8,9$ and choose the other five components of the gauge
field to be independent of the coordinates of the $S^4$ part. 

If a gauge field is switched on the physics on the D8-brane is
described by the Dirac--Born--Infeld action, eq.~\eqref{eq:dbi}. 
We are interested in trilinear Yukawa-like couplings containing
just one boson. The commutator in a non-Abelian field strength
necessarily introduces a second boson. This would only be relevant if
we would also consider two boson -- two fermion couplings. For this
reason there is no loss of generality by considering an Abelian gauge field.

\subsection{Trilinear Couplings}

Trilinear Yukawa-like couplings arise from terms of first order in the
field strength in the fermionic D8-brane action
\eqref{eq:fermionicaction}. 
Both $\Gamma_{D8}$ and $L_{D8}$ contain a piece linear in ${\cal F}$
\bea
  \Gamma_{D8}
  &=&\Gamma^{\underline{5}}(1-\frac{1}{2}\Gamma^{11}
  \Gamma^{\alpha\beta}{\cal F}_{\alpha\beta}+{\cal O}({\cal F}^2)),
  \\ 
   L_{D8}
  &=&-\Gamma^{\underline{5}} \Gamma^{11} \Gamma^{\alpha} 
   {\cal F}_{\alpha}^{\phantom{\alpha} \beta} D_\beta 
            + {\cal O}({\cal F}^2)
\eea
Inserting the expressions into the action, we obtain the Yukawa-like
coupling terms as 
\bea
S_3&=&i\frac{T}{2}\int \mathrm{d}^9\xi \sqrt{-g} \mathrm{e}^{-\phi}
\bar \Psi
\frac{1}{2}\Gamma_{\underline{5}}\Gamma^{11}\Gamma^{\alpha\beta}{\cal
  F}_{\alpha\beta} (\Gamma^\gamma D_\gamma-\Delta) \Psi \nn \\ 
&&-i\frac{T}{2}\int \mathrm{d}^9\xi \sqrt{-g} \mathrm{e}^{-\phi} \bar
\Psi
(1-\Gamma_{\underline{5}})\Gamma^{\underline{5}}\Gamma^{11}\Gamma^{\alpha}{\cal
  F}_{\alpha}^{\phantom{\alpha}\beta}D_\beta\Psi.  
\eea
This vertex is simplified by imposing the Majorana and Weyl condition,
\ie just keep terms with derivatives or three or seven $\Gamma$ matrices. 
The first piece just contributes
\bea
    S_3^{(1)}&=&-i\frac{T}{2}\int \mathrm{d}^9\xi \sqrt{-g} \mathrm{e}^{-\phi} 
    \bar \Psi \frac{1}{2}\Gamma_{\underline{5}}\Gamma^{\alpha\beta}{\cal F}_{\alpha\beta}\Gamma^\gamma \partial_\gamma\Psi
\eea
Decomposing the spinor in the same way as in the determination of the
spectrum eq.~\eqref{eq:c}, we see that this term vanishes.
Expanding the second piece gives the following
\bea
S_3^{(2)}&=&-i T
\int \mathrm{d}^9\xi \sqrt{-g} \mathrm{e}^{-\phi} 
\Bigg[ 
\bar \Psi  \frac{1}{2}\left(1-\Gamma^{\underline 5}\right) \nn\\
&&\frac{1}{\alpha^{3} V^{\frac{9}{4}}} \left(\Gamma^{\underline \mu}{\cal F}_{\mu\nu} \eta^{\nu\rho} \partial_\rho
+  M_{kk}^2 V^4 \Gamma^{\underline \mu}{\cal F}_{\mu w} \delta^{w w}\partial_w 
+  M_{kk}   V^2 \Gamma^{\underline w}{\cal F}_{w \mu} \eta^{\mu\rho} \partial_\rho
\right) \Psi\nn\\
  &&+\bar \Psi \frac{M_{kk}}{8\alpha^3} w~ V^{-13/4} {\cal F}_{\mu\nu} \Gamma^{\underline{ \mu\nu w}} \Psi\nn\\
  &&-\bar \Psi \frac{e^\phi}{16\alpha^2} F_{\underline {6789}} \left(  V^{-3/2} \Gamma^{\underline{\mu \nu}} {\cal F}_{\mu\nu} + 2 M_{kk} V^{1/2} \Gamma^{\underline{\mu w}} {\cal F}_{\mu w} \right) \Gamma^{\underline {56789}} \Psi\Bigg].
\eea
From now on, the raising and lowering of indices is understood to be
performed by $\eta_{\mu\nu}$ and $\delta_{w w}$. 
We also rescale $\Psi \to V^{-13/8} \Psi$.
This has no effect except changing the overall weighting function.
The term arising from the $w$ derivative can be dropped since it comes
with either one or two $\Gamma$ and does not contribute due to the
Majorana property of the spinors: 
\bea
  S_3 
  &=&-i\frac{\hat{T}}{\alpha^2 V_4 (2 \pi \alpha')^2} \int \mathrm{d}^4x \mathrm{d}w
  \mathrm{d}\Omega_4~  
  \Bigg[ \bar \Psi \frac{1}{2}\left(1-\Gamma^{\underline 5}\right)\nn\\
    &&
    \left(      V^{-\frac{7}{2}} \Gamma^{\underline \mu}{\cal
      F}_{\mu\nu} \partial^\nu 
    + M_{kk}^2 V^\frac{1}{2}    \Gamma^{\underline \mu}{\cal F}_{\mu
      w}  \partial^w 
    + M_{kk}   V^{-\frac{3}{2}} \Gamma^{\underline w}  {\cal F}_{w
      \mu}  \partial^\mu \right) \Psi \nn \\ 
    &&
    + \bar \Psi \frac{M_{kk}}{8} w~ V^{-\frac{9}{2}}  {\cal
      F}_{\mu\nu} \Gamma^{\underline{ \mu\nu w}}  \Psi\nn\\ 
    &&
    - \bar \Psi \frac{g_s\alpha^2 }{16}  F_{\underline {6789}} \left(
    V^{-2} \Gamma^{\underline{\mu \nu}}{\cal F}_{\mu\nu} + 2
    M_{kk}\Gamma^{\underline{\mu w}}{\cal F}_{\mu w} \right)
    \Gamma^{\underline {56789}}\Psi \Biggr]. 
\eea
We would also like to add a total derivative term to the interaction
\bea
S_3&=&\frac{i}{2}\frac{\hat{T} M_{kk}^2}{\alpha^2 V_4 (2 \pi \alpha')^2}  \int \mathrm{d}^4x \mathrm{d}w
  \mathrm{d}\Omega_4~\partial^w \left(\bar \Psi \frac{1}{2}\left(1-\Gamma^{\underline 5}\right)
   V^\frac{1}{2}    \Gamma^{\underline \mu}{\cal F}_{\mu w}   \Psi\right) \nn \\ 
&=&\frac{i}{2}\frac{\hat{T}M_{kk}^2 }{\alpha^2 V_4 (2 \pi \alpha')^2}  \int \mathrm{d}^4x \mathrm{d}w
  \mathrm{d}\Omega_4~\left(\partial^w\bar \Psi \frac{1}{2} \left(1-\Gamma^{\underline 5}\right)
   V^\frac{1}{2}    \Gamma^{\underline \mu}{\cal F}_{\mu w} \Psi \right. \nn \\
&&\qquad\qquad \qquad\qquad \qquad\qquad \left. + \bar \Psi \frac{1}{2} \left(1-\Gamma^{\underline 5}\right) V^\frac{1}{2}    \Gamma^{\underline \mu}{\cal F}_{\mu w}  \partial^w \Psi\right)\nn \\ 
\eea
in order to make the action more symmetric regarding the $w$ derivative. 
The terms where the derivative hits the function $V$ or $F_{\mu w}$ vanish due to the Majorana condition.

Now we employ the splitting~(\ref{eq:c}) we used for the spectrum,
perform the integration over the four-sphere and use the normalisation
for the $u$ and $\chi$ spinors
\bea
S_3
  &=&\!-i\frac{\hat T}{2\alpha^2} \int \mathrm{d}^4x \mathrm{d}w~ 
    \Bigg[ \bar \psi \left( V^{-7/2} \gamma^{\underline \mu}{\cal
      F}_{\mu\nu} \partial^\nu + M_{kk} V^{-3/2} \gamma {\cal F}_{w \mu} \partial^\mu \right.\nn\\
    && \left.
    + \frac{1}{2} M_{kk}^2 V^{1/2} \gamma^{\underline \mu}{\cal F}_{\mu w} \partial^w
    - \frac{1}{2} \overleftarrow{\partial^w} M_{kk}^2 V^{1/2} \gamma^{\underline \mu}{\cal F}_{\mu w} 
    \right)
    \psi \nn \\
    &&\!
    + \bar \psi \frac{M_{kk}}{4} w~ V^{-9/2}  {\cal F}_{\mu\nu}
    \gamma^{\underline{ \mu\nu}} \gamma \psi\nn\\ 
    &&\!
    -i \bar \psi \frac{M_{kk}}{4} \left(
    V^{-3} \gamma^{\underline{\mu \nu}}{\cal F}_{\mu\nu} + 2 M_{kk} V^{-1}
    \gamma^{\underline{\mu}} \gamma{\cal F}_{\mu w} \right) \psi
    \Biggr]. 
\eea 
We also re-express the flux
\begin{equation}
  F_{\underline{6789}}=\frac{3R^3}{g_s}
  \Bigl(\frac{3}{2}\frac{\alpha}{M_{kk}}V^{\frac{1}{4}}\Bigr)^{-4} 
  = \frac{2}{g_s}\frac{M_{kk}}{\alpha^2}V^{-1}.
\end{equation}

As next step we go to the Weyl basis according to eqs. (\ref{eq:weylspinor}) and (\ref{eq:weylbasis}) and expand the fermions and bosons in terms of complete sets. 
The effective four-dimensional action is then
\bea
   S_3 &=& M_{int}^{-2}\sum_{m,n,p} \int \mathrm{d}^4x \left[
     j_{m,n,p} F_{\mu\nu}^{(m)} \left( 
     \psi_+^{(n)\dagger} i\bar\sigma^{\mu} \partial^\nu \psi_+^{(p)} 
     + (-)^{k} 
     \psi_-^{(n)\dagger} i    \sigma^{\mu} \partial^\nu \psi_-^{(p)} 
     \right)\right.  \nn
     \\  
     &&\qquad \qquad 
     \left. +t_{m,n,p} M_{kk}^2 F_{\mu w}^{(m)}  \left( 
     \psi_+^{(n)\dagger} i\bar\sigma^{\mu} \psi_+^{(p)} 
     +(-)^{k} 
     \psi_-^{(n)\dagger} i    \sigma^{\mu} \psi_-^{(p)} 
     \right)\right.\nn 
     \\
     &&\qquad \qquad 
     \left. +l_{m,n,p} M_{kk} F_{\mu w}^{(m)}  \left( 
     \psi_+^{(n)\dagger} \partial^{\mu} \psi_-^{(p)} 
     + (-)^{k} 
     \psi_-^{(n)\dagger} \partial^{\mu} \psi_+^{(p)} 
     \right)\right. 
     \\
     &&\qquad \qquad 
     \left. +s_{m,n,p} M_{kk} F_{\mu\nu}^{(m)} \left(
     \psi_+^{(n)\dagger} \bar\sigma^{\mu\nu} \psi_-^{(p)} 
     + (-)^{k} 
     \psi_-^{(n)\dagger}     \sigma^{\mu\nu} \psi_+^{(p)} 
     \right)\right],\nn
\eea
where $k=m+n+p+1$ determines the parity properties of the effective vertex
and $M_{int}^{-2} \equiv 2 \pi \alpha' (\alpha \hat T)^{-1/2}/2$ is a dimensionful quantity setting the scale for the interactions.
The purely numerical coefficients are given by
\bea
  j_{m,n,p} &=& \int \mathrm{d}w~V^{-7/2} p_m(w) f_+^{n}(w) f_+^{p}(w), 
  \\ 
  t_{m,n,p} &=& \int \mathrm{d}w~\frac{1}{2} V^{1/2} p'_m \left( f_+^{n} f_+^{'p} -  f_+^{'n} f_+^{p}+ i V^{-3/2} f_+^{n} f_+^{p} \right),
  \\ 
  l_{m,n,p} &=& \int \mathrm{d}w~V^{-3/2} p'_m(w) f_+^{n}(w) f_-^{p}(w), 
  \\ 
  s_{m,n,p} &=& \int \mathrm{d}w~\frac{1}{4}\left(w V^{-9/2} + i V^{-3} \right) p_m(w) f_+^{n}(w) f_-^{p}(w).  
\eea
The coefficients have been simplified by using the parity properties
of the eigenfunctions $f_\pm^{(n)}$ eq.~(\ref{eq:plusminussymmetry})
and $p_m$ eq.~(\ref{eq:pparity}). 

Written in terms of the Dirac spinors \eqref{eq:mesinospinor}, this gives
\bea
   S_3 &=&-i M_{int}^{-2}\sum_{m,n,p} \int \mathrm{d}^4x \\
	&&\qquad
     \left[ F_{\mu\nu}^{(m)}
     \left( 
     j_{m,n,p}  \bar {\tilde{\psi}}^{(n)} \gamma^{\mu} (\gamma)^{k} \partial^\nu \tilde{\psi}^{(p)} 
     -s_{m,n,p}M_{kk} \bar {\tilde{\psi}}^{(n)} \gamma^{\mu\nu} (-\gamma)^{k}
     \tilde{\psi}^{(p)} \right)\right. \nn 
     \\  
     &&\qquad
     \left. + M_{kk} F_{\mu w}^{(m)}  \left( 
     l_{m,n,p} \bar{\tilde {\psi}}^{(n)} (-\gamma)^{k} \partial^{\mu} \tilde{\psi}^{(p)} 
     +t_{m,n,p}M_{kk} \bar{\tilde{\psi}}^{(n)} \gamma^{\mu} 
     (\gamma)^{k} \tilde \psi^{(p)}
     \right) \right].\nn
\eea
Since this action is effective, the non-renormalisability of the first three terms is not a problem.
The last term is a marginal term, which minimally couples the (axial) vector mesons to the (axial) 
vector current depending on the value of $k$.

Due to the properties
eq.~(\ref{eq:pparity}) and eq.~(\ref{eq:fparity})
the coupling constants satisfy the following symmetry properties
\begin{align}
	\nonumber
	j_{mpn} &= j_{mnp},
	& l_{mpn} &= -(-)^{k} l_{mnp},
	\\ \nonumber
        t^*_{mpn} &= -t_{mnp},
	& s^*_{mpn} &= -(-)^{k} s_{mnp}.
\end{align}

The integrals in these expressions can be computed numerically and
generically produce numbers of order 1.
Table~\ref{tab:coeffs} lists these numerical values for some
interactions. 
\begin{table}
\begin{center}
\begin{tabular}{l|r@{.}lr@{.}lr@{.}lr@{.}l}
  $m,n,p$ 
  & \multicolumn{2}{c}{$j_{m,n,p}$} 
  & \multicolumn{2}{c}{$t_{m,n,p}$} 
  & \multicolumn{2}{c}{$l_{m,n,p}$} 
  & \multicolumn{2}{c}{$s_{m,n,p}$} 
  \\ \hline
  $1,1,1$ & 0&42  & 0&$0+0.026i$ & 0&0 & 0&$0-0.11i$
  \\
  $1,1,2$ & 0&073 & 0&$029-0.064i$ & 0&11 & $-0$&$039-0.042i$
  \\
  $1,2,2$ & 0&26  & 0&$0-0.0061i$ & 0&0& 0&$0-0.051 i$
  \\
  $2,1,1$ & $-0$&068 & 0&$0+0.17i$ & $-0$&33 & $-0$&028
  \\
  $2,1,2$ & 0&16 & 0&$24+0.052i$ & 0&13 & 0&$011-0.049i$
  \\
  $2,2,2$ & 0&029 & 0&$0+0.056i$ & $-0$&040 & $-0$&041
\end{tabular}
\end{center}
\caption{\label{tab:coeffs}
  The numerical values of the coupling
  coefficients for the interactions of mesinos with (axial) vector mesons.} 
\end{table}
What is important for our purpose is not the exact values of these coefficients, 
but the fact that they are non-zero and not very small.

The value of the coupling constants are determined by these coefficients as well as by the factors $M_{int}$ and $M_{kk}$.
The string coupling $g_s$ and the constant $\alpha$ can be expressed in terms of gauge theory quantities $g_{YM}$ and $N_c$~\cite{Sakai:2004cn}:
\begin{equation}
 g_s = \frac{1}{2\pi}\frac{g_{YM}^2}{M_{kk}l_s};
 \qquad
 \alpha = \left(\frac{U_{kk}}{R}\right)^{\frac{3}{4}}
 = \frac{2}{3\sqrt{3}}(g_{YM}^2 N_c)^{\frac{1}{2}} M_{kk}l_s,
\end{equation}
in the regime $1\ll g_{YM}^2 N_c \ll g_{YM}^{-4}$~\cite{Kruczenski:2003uq}
giving
\begin{equation}
	\label{eq:mint}
 M_{int}^{4} = \frac{4}{3^9\pi^5} (g_{YM}^2N_c)^3 N_c M_{kk}^4.
\end{equation}
Notice that the only adjustable scale here is $M_{kk}$.
The coupling constants are therefore given by a fixed numerical factor times the appropriate power of $M_{kk}$.
Note that according to eq.~(\ref{eq:mint}) the interaction scale $M_{int}$ is proportional to the fourth root of the number of colours so that the interaction terms scale with $N_c^{-1/2}$.

To fix the numerical factors, we can use the condition that the rho meson mass and the pion decay constant should match experimental values.
The experimental rho meson mass is $776~\text{MeV}$, which gives $M_{kk}=949~\text{MeV}$.
The pion decay constant $f_{\pi}$ is related to the normalisation constant $C$ in eq.~(\ref{eq:p0norm}), giving
\begin{equation}
 f_\pi^2 = \frac{1}{27\pi^4} (g_{YM}^2 N_c) N_c M_{kk}^2.
\end{equation}
Using $N_c=3$ and 
matching this to the experimental value $f_{\pi}=92~\text{MeV}$ gives
$\lambda=g_{YM}^2 N_c = 8.2$.
With these values we find $M_{int}=173~\text{MeV}=0.18~M_{kk}$.
The 't~Hooft coupling satisfies the  requirement to be larger than one, which is necessary for using classical supergravity only. 
This restricts to small curvatures as usual in the AdS/CFT correspondence.
On the other hand, the requirement $\lambda \ll g_{YM}^{-4}$ is not met.
It arises from the requirement that the string coupling $g_s=\mathrm{e}^{\phi}$ ought to be small, which can only be satisfied up to a critical radius $U_{crit}$~\cite{Kruczenski:2003uq}. 
The main reason for this failure is that we are using $N_c=3$ while the $N_c\to \infty$ and $g_{YM} \to 0$ in the 't~Hooft limit.
If we go to the 't~Hooft limit with finite $\lambda$ the interaction scale goes off to infinity and the interactions are  suppressed.

Since the cross section for a particular interaction is proportional to the coupling constant and we have found the coupling constants to be non-zero and non-suppressed, we conclude that these interaction are not suppressed.

\label{sec:mes}
Other interesting quantities are the couplings of the 
mesinos to the external electromagnetic field and to the pion.  
We use the zero modes in the expansions eqs.(\ref{eq:fmn}) and (\ref{eq:fmw}).
The mesinos couple to the external electromagnetic gauge field through
\bea
  S_3 &=&-i M_{int}^{-2}\sum_{n,p} \int \mathrm{d}^4x \left[ \left( \partial_\mu \hat{\cal  V}_\nu(x)-\partial_\nu \hat{\cal V}_\mu(x) \right) \left( j_{{\cal V},n,p} \bar{\tilde\psi}^{(n)} \gamma^{\mu} (\gamma)^{n+p} \partial^\nu \tilde \psi^{(p)}\right.\right. \nn \\ 
  &&\qquad \qquad \qquad \qquad
     \left. \left. -s_{{\cal V},n,p} M_{kk}  \bar{\tilde\psi}^{(n)} \gamma^{\mu\nu} (-\gamma)^{n+p} \tilde\psi^{(p)} \right)\right],
\eea
with the coefficients given by
\bea
    j_{{\cal V},n,p} &=& 	     \int \mathrm{d}w~V^{-7/2} f_+^{(n)} f_+^{(p)}, \\
    s_{{\cal V},n,p} &=& \frac{1}{4} \int \mathrm{d}w~\left(w V^{-9/2} + i V^{-3} \right) f_+^{(n)} f_-^{(p)}.
\eea

The mesino interaction with the pion field is described by
\bea
  S_3 &=& \frac{M_{kk}}{f_\pi M_{int}^{2}} \sum_{n,p} \int \mathrm{d}^4x \left[ \partial_\mu \pi \left( M_{kk} t_{\pi,n,p} \bar{\tilde\psi}^{(n)} i\gamma^{\mu} (\gamma)^{n+p+1}\tilde \psi^{(p)}\right.\right. \nn \\
  &&\qquad \qquad \qquad \qquad \qquad
     \left. +l_{\pi,n,p}  \left. \bar{\tilde\psi}^{(n)} (-\gamma)^{n+p+1} \partial^{\mu} \tilde \psi^{(p)}\right)\right],
\eea
with the coefficients given by
\bea
    l_{\pi,n,p} &=&     	\int \mathrm{d}w~ V^{-9/2} f_+^{(n)} f_-^{(p)}, \\
    t_{\pi,n,p} &=& \frac{1}{2} \int \mathrm{d}w~ \left( f_+^{(n)} f_+^{'(p)} - f_+^{'(n)}f_+^{(p)} + i V^{-3/2} f_+^{(n)} f_+^{(p)} \right)V^{-5/2}.
\eea
The numerical values for the lowest mesino excitations are displayed in Table~\ref{tab:3}.
\begin{table}

\begin{center}
\begin{tabular}{l|r@{.}lr@{.}lr@{.}lr@{.}l}
  $n,p$ 
  & \multicolumn{2}{c}{$j_{{\cal  V},n,p}$} 
  & \multicolumn{2}{c}{$s_{{\cal  V},n,p}$} 
  & \multicolumn{2}{c}{$t_{\pi,n,p}$} 
  & \multicolumn{2}{c}{$l_{\pi,n,p}$} 
  \\ \hline
  $1,1$ & 0&79  & $-0$&$20i$ & 0&$21i$ & $-0$&38
  \\
  $1,2$ & 0&096 & 0&$089-0.080i$ & 0&$031i$ & 0&15
  \\
  $2,2$ & 0&58  & 0&$13i$ & 0&$14i$& $-0$&17
\end{tabular}
\end{center}
\caption{\label{tab:3} The numerical values of the coupling
  coefficients for the coupling of the mesinos to the electromagnetic field and the pion, respectively.} 
\end{table}
As in the case of the massive modes the couplings are less than unity but still significantly large such that they are not suppressed and have to be taken into account.
Mesinos couple in the Sakai--Sugimoto model to the photon and the pion. 
The couplings fall off as $N_c^{-1/2}$ and $1/N_c$ for the photon and for the pion, respectively.

\section{Discussion}

In this paper, we have investigated the fermionic sector of the flavour brane in the Sakai--Sugimoto model in the case of massless quarks and only one quark flavour, $\mathcal{N}_f=1$.
We found that there is a spectrum of mesinos (fermionic mesons) with masses on the same scale as the bosonic mesons, set by the compactification scale $M_{kk}$.
The presence of this fermionic sector itself is a problem for this model as a holographic descriptions of QCD as it is not part of QCD nor seen experimentally.
However, one could have hoped that the fermionic sector drops out in the low energy physics because of very high masses. 
Our result shows that this is not the case -- mesinos appear on the same energy scale as mesons.
This follows from the fact that the supersymmetry breaking scale is the same as the compactification scale or meson mass scale, $M_{kk}$.

Having found mesinos of similar mass as the mesons, we asked whether their interaction with the mesons is suppressed. 
If that had been the case, we could have argued that the fermionic mesinos represent harmless junk in the model, not affecting the meson physics. 
We found explicitly that this is not so for finite $N_c$. 
The interactions between the mesinos and the mesons are not suppressed and cannot be ignored.
The existence of mesinos that interact with the mesons is a serious problem for the Sakai--Sugimoto model.
This should in fact not come as a surprise, as there is only one energy scale, $M_{kk}$, in the model.

Most treatments of the D4/D8-brane model consider only the bosonic part of the probe brane action and therefore do not encounter the mesinos. 
If the aim with the holographic model simply were to reproduce QCD-like physics that would be justified. 
However, one of the main motivations for studying this model is that it is a superstring theory model where the D-branes are equipped with a supermultiplet of fields. 
We have in principle no choice but to include the fermionic sector of the D-brane probe action.

The explicit breaking of the supersymmetry due to boundary conditions on the D4-branes does not affect this point. 
The number of degrees of freedom in the particle spectrum is still that of a supersymmetric theory but the mass degeneracy is lifted as we saw in our calculation.

As said already, the problems exhibited in this paper stem from the fact that the supersymmetry breaking scale is the same as the meson mass scale. 
One solution to it could therefore be to disassociate these scales, \ie\ have a supersymmetry breaking scale different from $M_{kk}$. 
Then we could achieve a separation of mesino masses and meson masses.

\subsection*{Acknowledgements}
We are grateful to Hidehiko Shimada for useful discussions.
HGS is supported by the Alexander von Humboldt Foundation.

\bibliographystyle{JHEP}
\bibliography{references3}

\end{document}